\documentstyle[aps,prl,floats,twocolumn,epsfig]{revtex}
\begin{document}


\draft
\preprint{hep-ph/9711463\raisebox{7mm}{\hspace{-33mm}SNUTP 97-160}}
\twocolumn[\hsize\textwidth\columnwidth\hsize\csname @twocolumnfalse\endcsname
\title{Effective Field Theory for Low-Energy Two-Nucleon Systems}
\author{Tae-Sun Park$^{1}$, Kuniharu Kubodera$^{2}$, Dong-Pil Min$^{1}$
and Mannque Rho$^{3}$}
\address{
 ${}^1$Dept. of Physics and Center for Theoretical Physics,
 {\it Seoul Nat'l University, Seoul 151-742, Korea} \\
 ${}^2$Dept. of Physics and Astronomy,
 {\it University of South Carolina, Columbia, SC 29208, U.S.A.}\\
 ${}^3$Service de Physique Th\'{e}orique, CEA  Saclay,
\it 91191 Gif-sur-Yvette Cedex, France}
\date{December 9, 1997}
\maketitle

\begin{abstract}
We illustrate how effective field theories work in nuclear physics
by using an effective Lagrangian in which all other degrees of
freedom than the nucleonic one have been integrated out to
calculate
 the low-energy properties of two-nucleon systems, viz,
 the deuteron properties, the $np$ $^1S_0$ scattering
amplitude and the $M1$ transition amplitude entering into the
radiative $np$ capture process. Exploiting a finite cut-off
regularization procedure, we find all the two-nucleon low-energy
properties to be {\it accurately} described with little cut-off
dependence, in consistency with the general philosophy of effective
field theories.
\end{abstract}

\pacs{PACS numbers: 21.30.Fe, 13.75.Cs, 03.65.Nk, 25.40.Lw}
]

Effective field theories (EFTs) have long proven to be a powerful
tool in particle and condensed matter physics\cite{effective,pol}, so
it is quite natural that a considerable attention is nowadays paid
to the role of EFTs in nuclear physics where
phenomenological approaches have traditionally been tremendously
successful. Some authors have focused on nucleon-nucleon
interactions and two-nucleon
systems\cite{weinberg,pmr,vankolck,ksw,lm,cohen} while
some\cite{NRZ} on many-body systems including dense matter relevant
to  relativistic heavy-ion processes and compact stars. One of the
most spectacular cases was the recent chiral perturbation
calculation of the radiative $np$ capture at thermal energy
\cite{pmr} with an agreement with experiment within 1\%. What was
calculated in \cite{pmr} was however the meson-exchange current
corrections {\it relative} to the single-particle $M1$ matrix element
with the latter {\it borrowed} from the accurate Argonne $v_{18}$
\cite{v18} phenomenological two-nucleon wave function. In this
respect, one cannot say that it was  a complete calculation in the
framework of the given EFT, namely, chiral
perturbation theory (ChPT) although it was following the strategy
of \cite{weinberg} of using ChPT for computing ``irreducible
graphs" only.

The purpose of this Letter is to supply the ``missing link"  that
can render Ref.\cite{pmr} a ``first-principle calculation," that is to
 {\it obtain} the single-particle $M1$ matrix element within
the framework of EFTs \cite{friar}. In so doing
we will compute the static properties of the bound $np$ state (the
deuteron) and the $np$ scattering amplitude in the $^1S_0$ channel.
The results come out to be in a surprisingly good agreement with
the data, offering a first glimpse of how EFTs
work in nuclei.

Since we shall be interested in  very low-energy
processes with the energy scale $E\ll m_\pi\approx 140$ MeV, we
will integrate out all massive fields as well as the pion
field\cite{ksw}, leaving only the nucleon matter field
which can be treated in heavy-fermion formalism (HFF).
Since the anti-nucleon field also is integrated out in HFF,
there are no ``irreducible" loops 
(there will be, however, ``reducible" loops to
all orders in solving Lippman-Schwinger equation)
and the EFT becomes
non-relativistic quantum mechanics where all the interactions
appear in the potential.
Now the $np$ states we shall study are all very close
to the threshold: they are either weakly bound ($^3S_1$) or almost
bound ($^1S_0$). Bound states are not accessible by perturbation
expansion and the scattering state with a large scattering length
$a$ has a small scale $a^{-1}$, 
making the convergence of
EFTs highly non-trivial. We shall circumvent these difficulties by
summing ``reducible diagrams" -- which amounts to solving
Schr\"odinger (or Lippman-Schwinger) equation and using a cut-off
regularization  instead
of the usual dimensional regularization.

Due to the nonperturbative nature of the
Schr\"odinger equation,
unlike perturbative cases,
it does matter which regularization 
scheme one uses in effective theories.
Kaplan, Savage and Wise
\cite{ksw} and Luke and Manohar \cite{lm} have found that with the 
dimensional regularization, the EFT
breaks down at a very small scale, $p_{crit} = \sqrt{\frac{2}{a
r_e}}$ for large scattering length $a$, where $r_e$ is the
effective range and that this problem cannot be ameliorated by
introducing the pionic degree of freedom. As pointed out by 
Beane et al \cite{cohen} 
and Lepage\cite{lepage}, the problem can
however be resolved if one uses a cut-off regularization. In
effective theories, the cut-off has a physical meaning and hence it
should not be taken to infinity as one does in renormalizable
theories \cite{lepage}. In fact the strategy of effective field
theories is such that one should not pick either too low a cut-off
or too high a cut-off: if one chooses too low a cut-off, one risks
the danger of throwing away relevant degrees of freedom -- and hence
correct physics -- while if one chooses too high a cut-off, one
introduces irrelevant degrees of freedom and hence makes the theory
unnecessarily complicated. The astute in doing EFTs is in choosing
the proper cut-off. Thus with our effective Lagrangian in which the
lightest degree of freedom integrated out is the pion, the natural
cut-off scale is set by the pion mass. We find that the optimal
cut-off in our case is $\Lambda
\sim 200$ MeV as one can see from the results in Table 1 and Figures 1 and 2.

We shall do the calculation to the next-to-leading order (NLO).
The potential of the EFT is local and hence of zero range in coordinate
space, requiring regularization. 
In order to do the
calculation algebraically, we choose the following form of
regularization appropriate to a separable potential given by the
local Lagrangian:
\begin{equation}
\langle {\bf p'} | {\hat V} | {\bf p} \rangle
= S_\Lambda({\bf p}') V({\bf p}'-{\bf p}) S_\Lambda({\bf p})
\label{V}\end{equation}
where $S_\Lambda({\bf p}) = S(\frac{{\bf p}^2}{\Lambda^2})$ is the
regulator which suppresses the contributions from $|{\bf p}|
\gtrsim
\Lambda$, $\lim_{x\rightarrow 0} S(x) =1$ and $\lim_{x\gg
1}S(x)=0$, and $V({\bf q})$ is a finite-order polynomial in ${\bf
q}$. Up to the NLO, the most general form of $V({\bf q})$ is
\begin{equation}
V({\bf q}) = \frac{4\pi}{M} \left(C_0 +
(C_2 \delta^{ij} + D_2 \sigma^{ij}) q^i q^j \right),
\label{Vq}\end{equation}
where $M$ is the nucleon mass and
$\sigma^{ij}$ is the rank-two tensor that is effective only in
the spin-triplet channel,
\begin{equation}
\sigma^{ij} = \frac{3}{\sqrt{8}} \left(
\frac{\sigma_1^i \sigma_2^j + \sigma_1^j \sigma_2^i}{2}
- \frac{\delta^{ij}}{3} \sigma_1 \cdot \sigma_2 \right).
\end{equation}
Note that the coefficients $C_{0,2}$ are (spin) channel-dependent,
and that $D_2$ is effective only in spin-triplet channel.
Thus we have five parameters; two in $^1S_0$ and three in $^3S_1$ channel.
In principle, these parameters are calculable
from a fundamental Lagrangian (i.e., QCD) but nobody knows how to
do this. So in the spirit of EFTs, we shall fix them from
experiments. Since the explicit form of the regulator should not
matter\cite{lepage}, we shall choose the Gaussian form,
\begin{equation}
S_\Lambda({\bf p}) = \exp\left(- \frac{{\bf p}^2}{2\Lambda^2}\right)
\label{S}\end{equation}
where $\Lambda$ is the cut-off. (This form of the cut-off
functions, strictly speaking, upsets the chiral counting
\cite{cohen} on which we will have more to say later.) The
Lippman-Schwinger (LS) equation for the wavefunction
$|\psi\rangle$,
$
| \psi\rangle = | \varphi\rangle
+ {\hat G}^0 \,{\hat V} | \psi\rangle
$
where $|\varphi\rangle$ is the free wavefunction and ${\hat G}^0$
is the free two-nucleon propagator depending on the total energy
$E$, $
\langle {\bf p}' | {\hat G}^0 | {\bf p} \rangle
 = \frac{\langle {\bf p}' | {\bf p}\rangle}{
      E - \frac{{\bf p}^2}{M} + i 0^+}$
leads to
the $S$-wave function (for the potential (\ref{Vq})) of the form
\begin{eqnarray}
\psi({\bf r}) &=& \varphi({\bf r})
 + \frac{S(\frac{M E}{\Lambda^2})\,C_E}{1- \Gamma_E C_E} \,
  \left[
1 - \frac{\sqrt{Z} C_2}{C_E} (\nabla^2 + ME)
\right.
\nonumber \\
&&\left.
- \frac{\sqrt{Z} D_2}{C_E}
 \frac{S_{12}({\hat r})}{\sqrt{8}}
 r \frac{\partial}{\partial r} \frac{1}{r} \frac{\partial}{\partial r}\right]
 {\tilde \Gamma}_\Lambda({\bf r})
\label{fullwave}
\end{eqnarray}
where
\begin{eqnarray}
\Gamma_E &=&
4\pi \int \frac{d^3{\bf p}}{(2\pi)^3}
\frac{S_\Lambda^2({\bf p})}{ME - {\bf p}^2 + i0^+},
\label{GE}
\\
{\tilde \Gamma}_\Lambda({\bf r}) &=&
4\pi \int \frac{d^3{\bf p}}{(2\pi)^3} \frac{S_\Lambda({\bf p})}{ME - {\bf p}^2
 + i0^+} \mbox{e}^{i {\bf p}\cdot {\bf r}}\label{GLambda} ,
\\
Z &=& (1 - C_2 I_2)^{-2},
\label{Z}\\
C_E &=& a_\Lambda \left(1 + \frac12 a_\Lambda r_\Lambda ME\right)
 + (\sqrt{Z} D_2 ME)^2 \Gamma_E ,
\end{eqnarray}
with
\begin{eqnarray}
a_\Lambda &\equiv& Z\left[C_0 + (C_2^2 + \delta_{S,1} D_2^2) I_4)\right],
\\
r_\Lambda  &\equiv& \frac{2 Z}{a_\Lambda^2} \left[
   2 C_2 - (C_2^2 - \delta_{S,1} D_2^2) I_2 \right]
\label{arC}\end{eqnarray}
where $I_{n}$ ($n=2,\,4$) are defined by
\begin{equation}
I_{n} \equiv
 - \frac{\Lambda^{n+1}}{\pi} \int_{-\infty}^\infty
 dx\, x^{n} S^2(x^2).
\end{equation}
With the regulator (\ref{S}), the integrals come out to be $I_2=
-\frac{1}{2\sqrt{\pi}}
\Lambda^3$ and $I_4= -\frac{3}{4\sqrt{\pi}} \Lambda^5$.

The phase shifts can be calculated by looking at the large-$r$
behavior of the wavefunction. To do this, it is convenient to
separate the pole contributions of the integrals Eqs.(\ref{GE},
\ref{GLambda}) as
\begin{eqnarray}
\Gamma_E &=&  - i \sqrt{ME}\, S^2(\frac{ME}{\Lambda^2})
 + I_\Lambda(E),
\\
{\tilde \Gamma}_\Lambda({\bf r}) &=&
-\frac{S(\frac{ME}{\Lambda^2})}{r} \left[
 \mbox{e}^{i \sqrt{ME} r } - H(\Lambda r, \frac{ME}{\Lambda^2})
\right],
\end{eqnarray}
which define the functions $I_\Lambda(E)= \Lambda
I(\frac{ME}{\Lambda^2})$ and $H(\Lambda r, \frac{ME}{\Lambda^2})$,
both of which are real.  Note that $H(0,\,\varepsilon)=1$ which
makes ${\tilde \Gamma}_\Lambda({\bf 0})$ finite, and that
$\lim_{x\gg 1} H(x,\,\varepsilon)=0$. The phase shift $\delta$
takes the form
\begin{equation}
p\cot \delta = \frac{1}{S^2(\frac{ME}{\Lambda^2})} \left[
 I_\Lambda(E)
 - \frac{1-\eta^2(E)}{a_\Lambda (1 + \frac12 a_\Lambda r_\Lambda ME)}
\right],
\label{pCot1S0}\end{equation}
where the $\eta(E)$ is the $D/S$ ratio to be given below (see
(\ref{etaE})), which vanishes for the $^1S_0$ channel.

In order to fix the two coefficients $C_{0,2}$, we compare (\ref{pCot1S0}) to the {\it
effective-range expansion}
\begin{equation}
p\cot\delta = -\frac{1}{a} + \frac12 r_e p^2 + \cdots.
\label{ERT}
\end{equation}
We obtain
\begin{eqnarray}
\frac{1}{a_\Lambda} &=& \frac{1}{a} + \Lambda I(0)
 = \frac{1}{a} - \frac{\Lambda}{\sqrt{\pi}},
\\
r_\Lambda &=& r_e - \frac{2 I'(0)}{\Lambda} - \frac{4 S'(0)}{a \Lambda^2}
= r_e - \frac{4}{\sqrt{\pi}\Lambda} + \frac{2}{a \Lambda^2}.
\end{eqnarray}
These are essentially the ``renormalization conditions" in the
standard renormalization procedure. Two important observations to
make here: (a)  We note that there is an upper bound of $\Lambda$,
$\Lambda_{\rm Max}$, if one requires that $Z$ be positive and that
$C_2$ be real. That is, for $\Lambda
>\Lambda_{\rm Max}$, the potential of the EFT becomes
non-Hermitian. With $a= -23.732$ fm and $r_e= 2.697$ fm for the $^1S_0$
channel taken from the Argonne $v_{18}$ potential \cite{v18} (which
we take to be ``experimental"), we find that $\Lambda_{\rm Max}
\simeq 348.0$ MeV; (b)
the value $\Lambda_{Z=1}$ defined such that $Z=1$ when
$\Lambda=\Lambda_{Z=1} \simeq 172.2$ MeV is  quite special. At this
point, we have $r_\Lambda=0$ and $C_2=0$, that is, the NLO
contribution is identically zero. This corresponds to the
leading-order calculation with the $\Lambda$ chosen to fit the
experimental value of the effective range $r_e$.
A similar observation was made by Beane et al \cite{cohen} using
a square-well potential in coordinate space with a radius $R$, 
with $R^{-1}$ playing the role of $\Lambda$.

The resulting phase shift with $\Lambda=\Lambda_{Z=1}$ is plotted
in Fig.\ 1. We see that the agreement with the result taken from
the  Argonne $v_{18}$ potential \cite{v18} is perfect up to 
$p \sim 70$ MeV. Beyond that, we should expect corrections from
the next-to-next-order and higher-order terms. In Fig.\ 2, we show
how the phase-shift for a fixed center-of-mass momentum, 
$p= 68.5$ MeV varies as the cut-off is changed. The solid curve is
our NLO result, the dotted one the LO result (with $C_2=0$), and
the horizontal dashed line the result taken from the $v_{18}$
potential (``experimental"). We confirm that our NLO result is
remarkably insensitive to the value of $\Lambda$ for 
$\Lambda\gtrsim m_\pi$.

It is instructive to compare our result (\ref{pCot1S0}) with that
obtained with the dimensional regularization\cite{ksw},
\begin{equation}
\left. p\cot\delta\right|_{Dim.} = - a^{-1} (1 + \frac12 a r_e ME)^{-1}.
\label{dimension}
\end{equation}
Expanding $p\cot\delta$ of (\ref{dimension}) in $ME$, we find that
the coefficient of the $n$-th order term is order of $a^{n-1}
r_e^n$. This increases rapidly with $n$ when $a$ is large, disagreeing
strongly with the fact that the low-energy
scattering is well described by just two terms of the {\it
effective range expansion} in (\ref{ERT}). This observation led the
authors of \cite{ksw} to conclude that the critical momentum scale
at which the EFT expansion breaks down is very small for a very large
$a$:
\begin{equation}
\left. p_{crit}\right|_{Dim} \sim \sqrt{2/({a r_e})}.
\end{equation}
We arrive at a different conclusion. With the cut-off
regularization, the scattering length $a$ is replaced by an {\em
effective one}, $a_\Lambda$, that is order of $\frac{1}{\Lambda}$
for large $a$. This agrees with the findings of
Beane et al \cite{cohen} and Lepage \cite{lepage}.
Counting $r_e$ to be order of $\Lambda^{-1}$, the $n$-th order
coefficient now is $\Lambda^{1-2n}$, as one would expect on a
general ground.

\indent
The next quantity to consider is the transition $M1$ amplitude for
$n+p\rightarrow d+\gamma$ and the deuteron structure. For the $np$
capture, we need both the $^1S_0$ scattering wavefunction and the
deuteron wave function. The initial state wave function can be
written as
\begin{eqnarray}
\psi({\bf r}) &=&
\frac{\mbox{e}^{i \delta} \sin \delta}{ p r} u_0(r) | ^1S_0\rangle ,
\\
u_0(r) &=& \frac{\sin(pr + \delta)}{\sin\delta}
 - H(\Lambda r, \frac{ME}{\Lambda^2})
+ \beta_\Lambda {\cal D}(\Lambda r)
\end{eqnarray}
where $p=\sqrt{ME}$ is the center-of-mass momentum and
\begin{eqnarray}
{\cal D}(\Lambda r) &=&
 \frac{4\pi r}{\Lambda^2}
 \int \frac{d^3{\bf p}}{(2\pi)^3} S(\frac{{\bf p}^2}{\Lambda^2})
 \mbox{e}^{i {\bf p}\cdot {\bf r}},
\\
\beta_\Lambda &=& \frac{(\sqrt{Z}-1) \Lambda^2}{
 a_\Lambda(1 + \frac12 a_\Lambda r_\Lambda ME) I_2
 S(\frac{ME}{\Lambda^2})} .
\end{eqnarray}
\begin{figure}[t]
\centerline{\epsfig{file=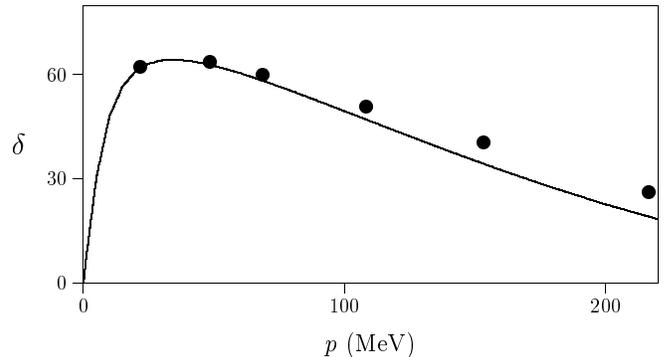,width=3.375in}}
\caption[phase]{\protect \small
$np$ $^1S_0$ phase shift (degrees) vs. 
the center-of-mass (CM) momentum $p$. 
Our theory with $\Lambda=\Lambda_{Z=1}\simeq
172$ MeV is given by the solid line, and the results from the
Argonne $v_{18}$ potential \cite{v18} (``experiments") by the solid
dots.}
\end{figure}

\begin{figure}[t]
\centerline{\epsfig{file=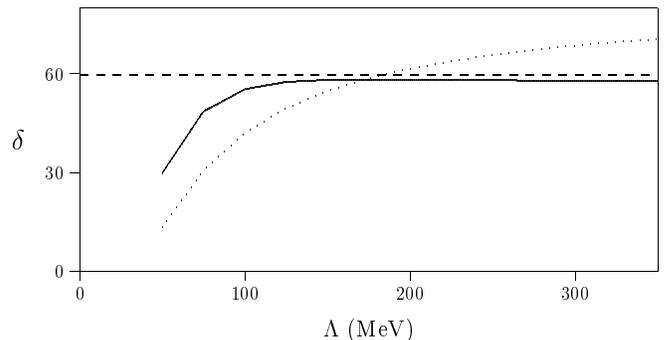,width=3.375in}}
\caption[deviate]{\protect \small
$np$ $^1 S_0$ phase shift (degrees) vs. the cut-off $\Lambda$ for a
fixed CM momentum $p= 68.5$ MeV. 
The solid curve represents the
NLO result, the dotted curve
 the LO result and the horizontal dashed line the
result from the $v_{18}$ potential \cite{v18}.}
\end{figure}
As for the $^3S_1$ coupled channel relevant for the final state of
the $np$ capture, 
we use the eigenphase parametrization \cite{eigenphase} with the
$\eta(E)$ given by $\eta(E)
\equiv -\tan \epsilon_1$ where $\epsilon_1$ is the mixing angle,
\begin{equation}
\frac{\eta(E)}{1- \eta^2(E)} =
\frac{\sqrt{Z} D_2  ME}{a_\Lambda\left(1 + \frac12 a_\Lambda r_\Lambda ME
\right)}
\label{etaE}.
\end{equation}
The $D_2$ so far undetermined can be fixed by the deuteron $D/S$
ratio $\eta_d\simeq 0.025$ \cite{v18} at $E=
-B_d$ with $B_d$ the binding energy of the deuteron,
\begin{equation}
\sqrt{Z} D_2 = \frac{\eta_d}{1-\eta_d^2}
 \frac{a_\Lambda}{-M B_d}
 \left[ 1 - \frac{1}{2} a_\Lambda r_\Lambda M B_d
 \right].
\end{equation}

Given $C_0$, $C_2$ and $D_2$ for a given $\Lambda$, all other
quantities are predictions. The binding energy of the deuteron is
determined by the pole position,
\begin{equation}
\gamma S^2(\frac{-\gamma^2}{\Lambda^2}) + I_\Lambda(-\gamma^2)
= \frac{1-\eta^2_d}{
  a_\Lambda \left(1 - \frac12 a_\Lambda r_\Lambda \gamma^2
 \right) }
\end{equation}
with $\gamma\equiv \sqrt{M B_d}$. The renormalization procedure is
the same as for the $^1S_0$ channel. The only difference is that
the value of $\Lambda_{Z=1}$ that makes $Z=1$ does not coincide
with $\Lambda_{r_\Lambda=0}$ that makes $r_\Lambda=0$. Using $a=
5.419$ fm and $r_e= 1.753$ fm \cite{v18} for the $^3S_1$ channel, we find
that $\Lambda_{\rm Max} = 304.0$ MeV, $\Lambda_{Z=1} = 198.8$ MeV and
$\Lambda_{r_\Lambda=0}= 216.1$ MeV. 
The resulting ($S$-wave and $D$-wave) radial
wavefunctions of the deuteron are
\begin{eqnarray}
u(r) &=& \mbox{e}^{-\gamma r} - H(\Lambda r, \frac{-\gamma^2}{\Lambda^2})
 + \beta_\Lambda {\cal D}(\Lambda r),
\\
\omega(r) &=& \eta_d \frac{r^2}{\gamma^2}
 \frac{\partial}{\partial r} \frac{1}{r} \frac{\partial}{\partial r}
 \frac{1}{r} \left[
 \mbox{e}^{-\gamma r} - H(\Lambda r, \frac{-\gamma^2}{\Lambda^2})
 \right].
\end{eqnarray}

We now have all the machinery to calculate the deuteron properties:
the wavefunction normalization factor $A_s$, the radius $r_d$, the
quadrupole moment $Q_d$ and the $D$-state probability $P_D$. The
magnetic moment of the deuteron $\mu_d$ is related to the $P_D$
through
\begin{equation}
\mu_d= \mu_S - \frac32 \left(\mu_S -\frac12\right) P_D
\label{mud}
\end{equation}
where $\mu_S \simeq 0.8798$ is the isoscalar
nucleon magnetic moment. Finally the one-body isovector $M1$
transition amplitude  relevant for $n+p \rightarrow d +
\gamma$ at threshold \cite{pmr} is
\begin{equation}
M_{\rm 1B}\equiv \int_0^\infty dr\, u(r) u_0(r).
\end{equation}

\begin{table}[t]
\caption[deuteronTable]{Deuteron properties and
the $M1$ transition amplitude entering into the $np$ capture for
various values of $\Lambda$.}\label{table1}
\begin{tabular}{ccccccc}
$\Lambda$ (MeV) & $150$ & $198.8$ & $216.1$ & $250$ & 
  Exp.\cite{v18} & $v_{18}$\cite{v18} \\
\hline
$B_d$ (MeV) & $1.799$ & $2.114$ & $2.211$ & $2.389$ & $2.225$ & 2.225\\
$A_s$ ($\mbox{fm}^{-\frac12}$)
   & 0.869 & 0.877 & 0.878 & 0.878 & 0.8846(8) & 0.885 \\
$r_d$ (fm)
   & 1.951 & 1.960 & 1.963 & 1.969 & 1.966(7) & 1.967 \\
$Q_d$ ($\mbox{fm}^2$)
   & 0.231 & 0.277 & 0.288 & 0.305 & 0.286 & 0.270 \\
$P_D$ (\%)
   & 2.11 & 4.61 & 5.89 & 9.09 & $-$ & 5.76 \\
$\mu_d$
   & 0.868 & 0.854 & 0.846 & 0.828 & 0.8574 & 0.847 \\
$M_{\rm 1B}$ (fm)
   & 4.06 & 4.01 & 3.99 & 3.96 & $-$ & 3.98
\end{tabular}
\end{table}

The (parameter-free) numerical results  are listed in Table~1 for
various values of the cut-off $\Lambda$. We see that the agreement
with the experiments (particularly for $\Lambda=216.1$ MeV)
is excellent with very little dependence on
the precise value of $\Lambda$. It may be coincidental but highly 
remarkable that even the quadrupole moment which as the authors of \cite{v18}
stressed, the $v_{18}$ potential fails to reproduce, comes out correctly.

%

We believe to have demonstrated the power of EFTs
in low-energy nuclear physics,
allowing us to be as close as one can hope to the
fundamental theory
in the sense put forward in Ref.\cite{effective,pol}. 
In particular, it
is satisfying that the classic $np$ capture process can be completely
understood from a ``first-principle" approach. Here the
cut-off regularization was found to be highly efficient: 
With the dimensional regularization 
the $M1$ matrix element was found to be in total disagreement with
the result of the Argonne $v_{18}$ potential. 

As mentioned, the Gaussian cut-off brings in terms higher order than NLO
which to be consistent, would require 
corresponding ``counter terms" in the potential although
our results indicate that the latter cannot be significant. The next task is to 
incorporate pions into the picture and go up in energy. This would enable 
us to explore the interplay between the breakdown 
of EFT and 
the emergence of a ``new physics", an important and generic
issue currently relevant in particle physics
where going beyond Standard Model is the Holy Grail. These 
issues will be
addressed in a forthcoming publication.
\section*{Acknowledgments}
\indent
The work of TSP and DPM was supported in part by the 
Korea Science and Engineering Foundation 
through CTP of SNU
and in part by the Korea Ministry of
Education under the grant (BSRI-97-2441, BSRI-97-2418) and the work of MR 
by a Franco-German Humboldt Research Prize and
Spain's IBERDROLA Visiting Professorship. KK is partially supported
by the NSF Grant No. PHYS-9602000.
TSP would like to thank H.K. Lee for his support
while he was in Hanyang University where part of this work was done.

\end{document}